%% file: main.tex
\documentclass[journal]{IEEEtran}
\usepackage{times}
\usepackage{epsfig}
\usepackage{graphicx}
\usepackage{amsmath}
\usepackage{subfigure}
\usepackage{algorithm}
\usepackage{algorithmic}
\usepackage{amssymb}
\usepackage{multicol}
\usepackage{multirow}
\usepackage{enumitem}
\usepackage{stfloats}
\usepackage{amsthm}
\usepackage{cite}
\usepackage{color}
\usepackage{relsize}
\usepackage{booktabs}
\usepackage{mathtools}
\usepackage{cuted}
\usepackage{tikz}
\usetikzlibrary{shapes, arrows, calc, positioning,matrix}
\tikzset{
data/.style={circle, draw, text centered, minimum height=3em ,minimum width = .5em, inner sep = 2pt},
empty/.style={circle, text centered, minimum height=3em ,minimum width = .5em, inner sep = 2pt},
}
\usepackage{environ}
\makeatletter
\newsavebox{\measure@tikzpicture}
\NewEnviron{scaletikzpicturetowidth}[1]{%
  \def\tikz@width{#1}%
  \begin{lrbox}{\measure@tikzpicture}%
  \BODY
  \end{lrbox}%
  \pgfmathparse{#1/\wd\measure@tikzpicture}%
  \BODY
}
\makeatother

\input{DefinedComand}

\usepackage{bm}
\DeclareMathAlphabet\mathbfcal{OMS}{cmsy}{b}{n}
\newcommand{\ten}[1]{\mathbfcal{#1}} %mathcal
\newcommand{\mat}[1]{\mathbf{#1}}

\usepackage[breaklinks=true,bookmarks=false]{hyperref}

\begin{document}

%%%%%%%%% TITLE
\title{Robust Factorization and Completion of Streaming Tensor Data via Variational Bayesian Inference}

\author{Cole Hawkins, Zheng Zhang,~\IEEEmembership{Member,~IEEE}
\thanks{C. Hawkins and Z. Zhang are with University of California Santa Barbara,
Santa Barbara, CA 93106, USA, (email: colehawkins@math.ucsb.edu, zhengzhang@ece.ucsb.edu)}}%
% \IEEEauthorblockN{Cole Hawkins}
% \IEEEauthorblockA{Department of Mathematics \\
% UC Santa Barbara,
% Santa Barbara, CA \\
% colehawkins@math.ucsb.edu}
% \and
% \IEEEauthorblockN{Zheng Zhang}
% \IEEEauthorblockA{Department of Electrical and Computer Engineering \\
% UC Santa Barbara, Santa Barbara, CA\\
% zhengzhang@ece.ucsb.edu}
% }

\maketitle
%\thispagestyle{empty}

%%%%%%%%% ABSTRACT
\begin{abstract}
Streaming tensor factorization is a powerful tool for processing high-volume and multi-way temporal  data in Internet networks, recommender systems and image/video data analysis. In many applications the full tensor is not known, but instead received in a slice-by-slice manner over time. Streaming factorizations aim to take advantage of  inherent temporal relationships in data analytics. Existing streaming tensor factorization algorithms rely on least-squares data fitting and they do not possess a mechanism for tensor rank determination. This leaves them  susceptible to outliers and vulnerable to over-fitting. This paper presents the first Bayesian robust streaming tensor factorization model. Our model successfully identifies sparse outliers, automatically determines the underlying tensor rank and accurately fits low-rank structure. We implement our model in Matlab and compare it to existing algorithms. Our algorithm is applied to factorize and complete various streaming tensors including synthetic data, dynamic MRI, video sequences, and Internet traffic data.
\end{abstract}
%%%%%%%%% BODY TEXT
\section{Introduction}

Multi-way data arrays (i.e., tensors) are collected in various application domains including recommender systems~\cite{karatzoglou2010multiverse}, computer vision~\cite{liu2013tensor}, medical imaging~\cite{adali2015multimodal}, chemometrics~\cite{morup2009automatic}, and uncertainty quantification~\cite{zhang2017big}. How to process, analyze and utilize such high-volume tensor data is a fundamental problem in machine learning, data mining and signal processing~\cite{sidiropoulos2017tensor,KoldaBader,morup2011applications,kolda2008scalable,anandkumar2014tensor}. Effective numerical techniques, such as CANDECOMP/PARAFAC (CP)~\cite{carroll1970analysis,harshman1970foundations}, Tucker~\cite{latent1}, and tensor-train~\cite{oseledets2011tensor} factorizations, have been proposed to compress full tensors and to obtain their low-rank representations. Extensive optimization and statistical techniques have also been developed to obtain the low-rank factors and to predict the full tensor of an incomplete (and possibly noisy) multi-way data array~\cite{zhou2015bayesian,cpOptCompletion,Riemannian,SNN}. The process of recovering a full tensor based on its complete samples is called {\it tensor completion}.

This paper is interested in the factorization and completion of streaming tensors. Streaming tensors are multi-way data arrays that appear sequentially in the time domain. Incorporating temporal relationships in tensor data analysis can give significant advantages, and such techniques have been applied in anomaly detection~\cite{fanaee2016tensor}, discussion tracking~\cite{bader2008discussion} and context-aware recommender systems~\cite{tensor_recommender_systems}. In such applications  the current temporal relationships are of  high interest. By computing factorizations in a streamed manner one avoids both irrelevant information and the computational overhead associated with long-past data. A large body of low-rank streaming data analysis can be traced back to the projection approximate subspace tracking \cite{PAST}, which address two-way data. In the past decade, streaming tensor factorization has been studied under several low-rank tensor models, such as the Tucker model in \cite{sun2006beyond} and the CP decomposition in \cite{shaden,online-cp,online-SGD,kasai2016online}. These approaches are similar in the sense of choosing their objective functions, but differ in choosing their specific numerical optimization solvers. For instance, least-square optimization is used in \cite{kasai2016online} and stochastic gradient descent is employed in \cite{online-SGD}. All existing streaming tensor factorizations assume a fixed rank, but it is hard to estimate the rank {\it a priori}. Additionally, no existing techniques  can capture the sparse outliers in a streaming tensor, although some techniques have been proposed for non-streaming data~\cite{wright2009robust,lu2016tensor,goldfarb2014robust,huang2015provable,zhao2015bayesian, zhao2016bayesian}.

{\bf Paper Contributions.} This paper proposes a new method for the {\em robust factorization and completion} of streaming tensors. Here ``robustness'' means the ability to capture sparsely corrupted data or outliers. This can be employed in many applications such as dynamic MRI~\cite{otazo2015low} and network anomaly detection~\cite{li2011robust}. We model the whole temporal tensor dataset as the sum of a low-rank streaming tensor and a time-varying sparse component. In order to capture these two different components, we present a Bayesian statistical model to enforce low-rank and sparsity via hyper-parameters and proper prior density functions. The posterior probability density function (PDF) of the hidden factors is then computed by the variational Bayesian method~\cite{winn2005variational}. The variational Bayesian method was previously employed in~\cite{babacan2012sparse} and~\cite{zhao2015bayesian,zhao2016bayesian} to solve non-streaming low-rank matrix and tensor completion problems, respectively. Therefore, our work can can be regarded as an extension of~\cite{babacan2012sparse,zhao2015bayesian,zhao2016bayesian} to streaming tensors with sparse outliers. Since robust streaming tensor factorization is very different from standard tensor factorization, our proposed probabilistic model and the variational Bayesian solver also differ remarkably from those in~\cite{babacan2012sparse,zhao2015bayesian,zhao2016bayesian}.

\section{Preliminaries and Notations}

%\subsection{Notations}
%We will review the relevant tensor background. For a more thorough exposition, see \cite{KoldaBader}.

Throughout this paper, we use a bold lowercase letter (e.g., $\mat{a}$) to represent a vector, a bold uppercase letter (e.g., $\mat{A}$) to represent a matrix, and a bold calligraphic letter (e.g., $\ten{A}$) to represent a tensor. A tensor is a generalization of a matrix, or a multi-way data array. More formally, an order-$N$ tensor is a $N$-way data array $\ten{A}\in \mathbb{R}^{I_1 \times I_2 \times \dots \times I_N}$, where $I_k$ is the size of mode $k$. Given the integer $i_k \in [1, I_k] $ for each mode $k=1 \cdots N$, an entry of the tensor $\ten{A}$ is denoted as $a_{i_1,\cdots, i_N}$.

\begin{definition}
Let $\ten{A}$ and $\ten{B}$ be two tensors of the same dimensions, then their {\bf inner product} is defined as
\begin{equation}
\langle \ten{A},\ten{B} \rangle = \sum_{i_1=1}^{I_1} \dots \sum_{i_N=1}^{I_N}a_{i_1,\dots,i_N}b_{i_1,\dots,i_N}.\nonumber
\end{equation}
\end{definition}
Based on tensor inner product, the {\bf Frobenius norm} of tensor $\ten{A}$ is defined as
\begin{equation}
\label{eq: tensor norm}
||\ten{A}||_{\rm F} = \langle \ten{A},\ten{A} \rangle^{1/2}.
\end{equation}

% It is often expensive to store all elements of a large-scale multi-way tensor. Fortunately, many realistic tensors have a low-rank property. This allows us to significantly compress the data and perform fast tensor computation. In this paper, we will use the CANDECOMP/PARAFAC (CP) factorization~\cite{carroll1970analysis,harshman1970foundations}.
\begin{definition}
A $N$-way tensor $\ten{T} \in \mathbb{R}^{I_1\times \cdots \times I_N}$ is \textbf{rank-1} if it can be written as a single outer product of $N$ vectors
\begin{equation}
\ten{T} =\mat{a}^{(1)} \circ \dots \circ \mat{a}^{(N)}, \; \text{with} \; \mat{a}^{(k)} \in\mathbb{R}^{I_k} \; \text{for}\; k=1,\cdots, N. \nonumber
\end{equation}
\end{definition}

%The constant $\sigma$ is not strictly necessary, but it allows us to require that henceforth all rank-1 factors have norm one.
\begin{definition}\label{Def: CP}
The {\bf CP factorization}~\cite{carroll1970analysis,harshman1970foundations} expresses a $N$-way tensor $\ten{A}$ as the sum of multiple rank-1 tensors:
\begin{equation}
\ten{A} = \sum_{r=1}^Rs_r \mat{a}_r^{(1)} \circ \dots \circ \mat{a}_r^{(N)}, \; \text{with} \; \mat{a}_r^{(k)} \in\mathbb{R}^{I_k}.
\end{equation}
Here the minimal integer $R$ that ensures the equality is called the {\bf CP rank} of $\ten{A}$. The determination of a CP rank is NP-hard \cite{CPRankNPHard}, therefore in practice one relies on numerical techniques to provide a good approximation.
\end{definition}

For convenience, we express the CP factorization as
\[
\ten{A} = \sum_{r=1}^R s_r\mat{a}_r^{(1)} \circ \dots \circ \mat{a}_r^{(N)} = [\![ \mat{A}^{(1)},\dots,\mat{A}^{(N)};\mat{s}]\!],
\]
where the $\{ \mat{a}_r^k\}_{r=1}^R$ form the columns of the matrix $\mat{A}^{(k)}$. It is convenient to express this matrix both column-wise and row-wise, so we include two means of expressing a factor matrix
\[
\mat{A}^{(k)} = [\mat{a}_1^{(k)},\dots ,\mat{a}_R^{(k)}]=[\hat{\mat{a}}_1^{(k)};\dots ;\hat{\mat{a}}_{I_k}^{(k)}] \in \mathbb{R}^{I_k\times R}.
\]
%We will never mix row-wise and column-wise expressions, so the meaning of a vector $\mat{a}$ will be clear from context.
Here $\mat{a}_j^{(k)}$ and $\hat{\mat{a}}_{i_k}^{(k)}$ denote the $j$-th column and $i_k$-th row of $\mat{A}^{(k)}$, respectively. We will primarily use the {column}-wise expression, but the {row}-wise definition provides more a more intuitive presentation in our subsequent Bayesian model.

\begin{definition}
The {\bf generalized inner product} of $N$ vectors of the same dimension $I$ is defined as
\[
\langle \mat{a}^{(1)},\dots,\mat{a}^{(N)} \rangle = \sum\limits_{i=1}^I\prod\limits_{k=1}^N a^{(k)}_i.
\]
\end{definition}

We can now express the entries of a low-rank tensor $\ten{A}$ as in Definition \ref{Def: CP} by a generalized inner product of the rows of the factor matrices.
\[
a_{i_1,\dots,i_N} = \langle \hat{\mat{a}}_{i_1}^{(1)},\dots,\hat{\mat{a}}_{i_N}^{(N)} \rangle.
\]

%We fit entire factor matrices to the data, so most of our computations will be performed on matrices that encode the low-rank factors. Therefore we require two matrix operations. The first will be useful in computing expectation values in our Bayesian model.
\begin{definition}
The {\bf Hadamard product} of two matrices of the same dimensions is the entry-wise product and is written $\mat{A}\circledast \mat{B}$. This is extended to $N$ matrices $\{\mat{A}^{(n)}\}$ in the natural manner and is written
\[
\hadamard_n \mat{A}^{(n)} = \mat{A}^{(1)}\circledast\mat{A}^{(2)}\circledast\cdots\circledast\mat{A}^{(N)}.
\]
\end{definition}
We will need to construct a low-rank tensor from the factor matrices, so we introduce a corresponding matrix product.
\begin{definition}
The {\bf Khatri-Rao product} of two matrices $\mat{A} \in \mathbb{R}^{I\times R}$ and $\mat{B} \in \mathbb{R}^{J\times R}$ is the columnwise Kronecker product, and is written as
\[
\mat{A}\odot\mat{B}= [\mat{a}_1 \otimes \mat{b}_1, \ldots, \mat{a}_R \otimes \mat{b}_R] \in \mathbb{R}^{IJ\times R}.
\]
We will use the product notation to denote the Khatri-Rao product of $N$ matrices in reverse order:
\[
\bigodot_n \mat{A}^{(n)} = \mat{A}^{(N)}\odot \mat{A}^{(N-1)}\odot\cdots\odot\mat{A}^{(1)}.
\]
If we exclude the $k$-th factor matrix, the Khatri-Rao product can be written as
\[
\bigodot_{n\neq k} \mat{A}^{(n)} = \mat{A}^{(N)}\odot\cdots\odot \mat{A}^{(k+1)}\odot\mat{A}^{(k-1)}\odot\mat{A}^{(1)}.
\]
\end{definition}

\section{Review of Streaming Tensor Factorization}
\begin{figure}[t]
\begin{center}
\includegraphics[width=3in]{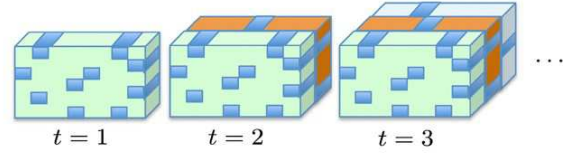}
\caption{A stream of partially observed tensors, adopted from Fig. 1 in\cite{online-SGD}.}
\label{fig:tensor_stream}
\end{center}
\end{figure}

Let $\{\ten{X}_t\}$ be a temporal sequence of $N$-way tensors, where $t \in \mathbb{N}$ is the time index and $\ten{X}_t$ of size $I_1\times \dots \times I_N$ is a slice of this multi-way stream. Streaming tensor factorizations aim to extract the latent tensor factors evolving with time. In this paper, we consider the CP factors of streaming tensors.

In order to compute a rank-$R$ streaming factorization, at each time point $t=T$ one can consider all slices from $t=i$ to $t=T$, a window size of $w=T-i+1$. One can seek for $N$ non-temporal factor matrices $\{\mat{A}^{(k)}\in\mathbb{R}^{I_k\times R}\}_{k=1}^N$  and a temporal factor matrix $\mat{A}^{(N+1)}\in\mathbb{R}^{(T-i+1)\times R}$ to approximate all multi-way data in this time window. The standard formulation for streaming tensor factorization is given below~\cite{online-SGD}:
\begin{equation}
\label{eq: basic streaming}
\min_{\{\mat{A}^{(k)}\}_{k=1}^{N+1}} \sum_{t=i}^T \mu^{T-t} \|\ten{X}_t - [\![ \mat{A}^{(1)},\dots,\mat{A}^{(N)};\hat{\mat{a}}^{(N+1)}_{t-i+1}]\!]\|_F^2.
\end{equation}
The parameter $\mu\in(0,1)$ is a forgetting factor, and $\{\mat{A}^{(i)}\}$ are the discovered CP factors. Please note that $\hat{\mat{a}}^{(N+1)}_{t-i+1}$ denotes one row of the temporal factor matrix $\mat{A}^{(N+1)}$. The exponentially weighted forgetting factor controls the weight of the past data, and the sliding window size $T-i+1$ can be specified by the user based on the available computing and memory resources.

In many real applications, only partial data $\ten{X}_{t,\Omega_t}$  is observed at each time point (see Fig. \ref{fig:tensor_stream}). Here $\Omega_t$ denotes the index set of the partially observed entries. For a general $N$-way tensor $\ten{X}$ and a sampling set $\Omega$, we have
\begin{equation}
\ten{X} _{\Omega }=\left\{\begin{matrix}
x_{i_1,\cdots,i_N}& \; {\rm if} \; {(i_1,i_2,\cdots,i_N)} \in \Omega\\ 
0 & \; {\rm otherwise}. \;\;\;\;\;\;\;\;\;\;
\end{matrix}\right. \nonumber
\end{equation}
For notational convenience we will compress subscript $(t,\Omega_t)$ to $\Omega_t$ so 
\[
\ten{X}_{\Omega_t}:=\ten{X}_{t,\Omega_t}.
\]
In the presence of missing data the underlying hidden factors can be computed to impute missing entries by solving the following {\it streaming tensor completion} problem:
\begin{equation}
\label{eq: streaming completion}
\min_{\{\mat{A}^{(k)}\}_{k=1}^{N+1}} \sum_{t=i}^T \mu^{T-t} \|\left(\ten{X}_t - [\![ \mat{A}^{(1)},\dots,\mat{A}^{(N)};\hat{\mat{a}}^{(N+1)}_{t-i+1} ]\!]\right)_{\Omega_t} \|_{\rm F}^2.
\end{equation}

 % and draw conclusions about the data stream
Existing streaming factorization and completion frameworks~\cite{online-cp,online-SGD,kasai2016online} solve \eqref{eq: basic streaming} and \eqref{eq: streaming completion}  as follows: at each time step one updates the $N$ non-temporal factor matrices $\mat{A}^{(j)}\in\mathbb{R}^{I_j\times R}$ and  $\{\hat{\mat{a}}^{(N+1)}_{t-i+1}\}$.
%the final row of the ${(N+1)}^{st}$ factor matrix
By fixing the past time factors, these approaches provide an efficient updating scheme to solve the above non-convex problems.

\section{Bayesian Model for Robust Streaming Tensor Factorization \& Completion}

In this section, we present a Bayesian method for the robust factorization and completion of streaming tensors $\{ \ten{X}_t\}$.

\subsection{An Optimization Perspective}
In order to simultaneously capture the sparse outliers and the underlying low-rank structure of a streaming tensor, we assume that each tensor slice $\ten{X}_t$ can be fit by
\begin{equation}\label{eq:slice L+S}
\ten{X}_t = \tilde{\ten{X}}_t+\ten{S}_t+\ten{E}_t.
\end{equation}
Here $\tilde{\ten{X}}_t$ is low-rank, $\ten{S}_t$ contains sparse outliers, and $\ten{E}_t$ denotes dense noise with small magnitudes. The low-rank and sparse components are of independent interest. For example, in recommender systems the low-rank structure should inform recommendations, and sparse outliers may be flawed ratings that are best ignored. In network traffic, the low-rank component can inform an administrator of the usual traffic flow while sparse outliers indicate anomalies that should be investigated.

Assume that each slice $\ten{X}_t$ is partially observed according to a sampling index set $\Omega_t$. Note that the sampling set can be different as time evolves. Based on the partial observations $\{ \ten{X}_{\Omega_t}\}$, we will solve a streaming tensor completion problem to find a reasonable low-rank factors for $\{\tilde{\ten{X}}_t\}$ in the specified time window $t \in [T-i+1, T]$ as well as the sparse component $\ten{S}_t$. This problem simplifies to robust streaming tensor factorization if $\Omega_t$ includes all possible indices, in other words, the whole tensor slice is given at every time step.

In order to enforce the low-rank property of $\tilde{\ten{X}}_t \in \mathbb{R}^{I_1\times\cdots \times I_N}$, we assume the following CP representation in the time window $t \in [i, T]$:
\begin{equation}
\tilde{\ten{X}}_t=[\![ \mat{A}^{(1)},\dots,\mat{A}^{(N)};\hat{\mat{a}}^{(N+1)}_{t-i+1}]\!]. \nonumber
\end{equation}
The sparsity of $\ten{S}_t$ can be achieved by enforcing its 1-norm $\| \ten{S}_t\|_1$ to be small. Therefore, by modifying \eqref{eq: streaming completion}, we have the following optimization problem:

\begin{align}
\label{eq: sparse completion streaming} 
 \nonumber\min_{\{\mat{A}^{(j)}\},\ten{S}_{\Omega_T}} \ &\sum_{t=i}^{T-1} \mu^{T-t} \|\left(\tilde{\ten{D}}_t  - [\![\mat{A}^{(1)},\dots,\mat{A}^{(N)};\hat{\mat{a}}^{(N+1)}_{t-i+1} ]\!]\right)_{\Omega_t}\|_{\rm F}^2\\ 
 \nonumber +&\|\ten{Y}_{\Omega_T}-\ten{S}_{\Omega_T}-\left([\![\mat{A}^{(1)},\dots,\mat{A}^{(N)};\hat{\mat{a}}^{(N+1)}_{T-i+1} ]\!]\right)_{\Omega_T}\|_{\rm F}^2\\
 +&\alpha\|\ten{S}_{\Omega_T}\|_1.
\end{align}
In our notation $\ten{Y}_{\Omega_T}=\ten{X}_{T,\Omega_T}$ is the observation of current slice, $\ten{S}_{\Omega_T}$ is its outliers, and $\{\tilde{\ten{D}}_{\Omega_t}\}_{t=i}^{T-1}$ are the observed past slices {\em with their sparse errors removed}. Once the robust completion or factorization of all previous slices is done, $\tilde{\ten{D}}_t$ can be obtained as $\tilde{\ten{D}}_t=\ten{X}_t -\ten{S}_t$.

One of the key challenge in solving (\ref{eq: sparse completion streaming}) is the determination of the rank $R$. If the rank is too large the computation will be expensive and the model will over-fit. If the rank is too small the model will not capture the full data structure. It is also non-trivial to select a proper regularization parameter $\alpha$. In order to fix these issues, we develop a Bayesian model which can automatically determine these parameters.

\subsection{Probabilistic Model for (\ref{eq:slice L+S}) }
{\bf Likelihood:} We first need to define a likelihood function for the data  $\ten{Y}_{\Omega_T}$ and $\{\tilde{\ten{D}}_{\Omega_t}\}_{t=i}^{T-1}$ based on (\ref{eq:slice L+S}) and \eqref{eq: sparse completion streaming}. We discount the past observations outside of the time window. We also use the forgetting factor $\mu< 1$ to exponentially weight the variance terms of past observations. This permits long-past observations to deviate significantly from the current CP factors with little impact on the current CP factors. Therefore, at time point $t=[i,T]$, we assume that the Gaussian noise has a 0 mean and variance $(\mu^{T-t} \tau)^{-1}$. This leads to the likelihood function in (\ref{eq:observationModel}). In this likelihood function, $\tau$ specifies the noise precision, $\hat{\mat{a}}_{i_n}^{(n)}$ denotes the $i_n$-th row of $\mat{A}^{(n)}$, and $\ten{S}_{\Omega_T}$ only has values corresponding to observed locations.
\begin{figure*}[t]
%\small
\begin{align}
\label{eq:observationModel}
p\left(\ten{Y}_{\Omega_T},\{\tilde{\ten{D}}_{\Omega_t}\}\middle\vert\{\mat{A}^{(n)}\}_{n=1}^{N+1},\ten{S}_{\Omega_T},\tau\right) = & \prod_{(i_1,\dots,i_n)\in \Omega_T}
\mathcal{N}\left(\ten{Y}_{i_1 \ldots i_N} \middle\vert \left\langle\hat{\mat{a}}^{(1)}_{i_1},\cdots,\hat{\mat{a}}^{(N)}_{i_N},\hat{\mat{a}}^{(N+1)}_{T-i+1}\right\rangle + \mathcal{S}_{i_1 \ldots i_N}, \tau^{-1}\right) \times \nonumber \\
&\prod_{t=i}^{T-1}\prod_{(i_1,\dots,i_n)\in\Omega_t}\mathcal{N}\left({\tilde{\ten{D}}}_{t,{i_1 \ldots i_N}} \middle\vert \left\langle\hat{\mat{a}}^{(1)}_{i_1},\cdots,\hat{\mat{a}}^{(N)}_{i_N},\hat{\mat{a}}^{(N+1)}_{t-i+1},\right\rangle, (\tau\mu^{T-t})^{-1}\right).
\end{align}
\normalsize
\end{figure*}

In order to infer the unknown factors and sparse terms in our streaming tensor factorization/completion, we should also specify their prior distributions. 

{\bf Prior Distribution of $\{ \mat{A}^{(n)}\}$:} We assume that each row of $\mat{A}^{(n)}$ obeys a Gaussian distribution and that different rows are independent to each other. Similar to~\cite{zhao2015bayesian}, we define the prior distribution of each factor matrix as
\begin{equation}
\label{eq:priorA}
p\big(\mat{A}^{(n)}\big\vert \boldsymbol\lambda \big) = \prod_{i_n=1}^{I_n} \mathcal{N}\big(\hat{\mat{a}}_{i_n}^{(n)} \big\vert \mathbf{0}, \boldsymbol\Lambda^{-1} \big), \, \forall n\in [1,N+1] 
\end{equation}
where $\boldsymbol\Lambda=\text{diag}(\boldsymbol\lambda)\in\mathbb{R}^{R\times R}$ denotes the precision matrix. All factor matrices share the same covariance matrix. Note that the $r$-th column of all factor matrices share the same precision parameter $\lambda_r$, and a large $\lambda_r$ will make the $r$-th rank-1 term more likely to have a very small magnitude. Therefore, by controlling the hyper parameters $\boldsymbol\lambda \in \mathbb{R}^R$ , we can tune the rank of our CP model. This process will be specified in Section~\ref{subsec: hyper}.

{\bf Prior Distribution of $\ten{S}_{{\Omega}_T}$:} Similar to the low-rank factors, we also place a Gaussian prior distribution over the component $\ten{S}_{ \Omega_T}$:
\begin{equation}
\label{eq:priorS}
p(\ten{S}_{\Omega_T}\vert\boldsymbol\gamma) = \prod_{(i_1,\ldots,i_N) \in \Omega_T}\mathcal{N}(\ten{S}_{i_1 \ldots i_N}\vert 0, \gamma_{i_1 \ldots i_N}^{-1}),
\end{equation}
where $\boldsymbol{\gamma}$ denotes the sparsity precision parameters. If $\gamma_{i_1 \ldots i_N}$ is very large, then the associated element in $\ten{S}_{\Omega_T}$ is likely to have a very small magnitude. By controlling the value of $\gamma_{i_1 \ldots i_N}^{-1}$, we can control the sparsity of $\ten{S}_{\Omega_T}$. The process of determining $\gamma_{i_1 \ldots i_N}^{-1}$ will also be discussed in Section~\ref{subsec: hyper}.
\begin{figure*}
\begin{equation}\label{eq:posterior density}
p\left(\Theta\middle\vert\ten{Y}_{\Omega_T},\{\tilde{\ten{D}}_{\Omega_t}\}\right) =
\frac{p\left(\ten{Y}_{\Omega_T},\{\tilde{\ten{D}}_{\Omega_t}\}\middle\vert\ \{\mat{A}^{(n)}\}_{n=1}^{N+1},\ten{S}_{\Omega_T},\tau \right)
\left\{
{\prod\limits_{n=1}^{(N+1)}p\big(\mat{A}^{(n)}}\big\vert \boldsymbol\lambda \big) \right\} p(\boldsymbol\lambda)
p(\ten{S}_{\Omega_T}\vert\boldsymbol\gamma)p(\boldsymbol\gamma)p(\tau)}{p(\ten{Y}_{\Omega_T},\{\tilde{\ten{D}}_{\Omega_t}\})}.
\end{equation}
\hrulefill
% The spacer can be tweaked to stop underfull vboxes.
\vspace*{4pt}
\end{figure*}

\subsection{Prior Distribution of Hyper Parameters}
\label{subsec: hyper}

We still have to specify three groups of hyper parameters: $\tau$ controlling the noise term, $\boldsymbol \lambda$ controlling the CP rank, and $\{ \gamma_{i_1 \ldots i_N}\}$ controlling the sparsity of $\ten{S}_{\Omega_T}$. Instead of assigning them deterministic values, we treat them as random variables and assign them Gamma prior distributions:
\begin{equation}
\label{eq:prior_hyper}
\begin{split}
p(\tau) &= \text{Ga}(\tau\:\vert \: a_0^\tau, b_0^\tau),\\
p(\boldsymbol\lambda) &= \prod_{r=1}^{R}\text{Ga}(\lambda_r \vert c_0, d_0),\\
p(\boldsymbol\gamma) &= \prod_{(i_1,\ldots,i_N)\in\Omega_T}\text{Ga}(\gamma_{i_1 \ldots i_N }\vert a_0^\gamma, b_0^\gamma).
\end{split}
\end{equation}
A Gamma distribution can be written as
\[
\text{Ga}(x\vert a,b) = \frac{b^a x^{a-1} e^{-bx}}{\Gamma(a)}, \]
where $\Gamma(a)$ is the Gamma function. The Gamma distribution provides a good model for our hyper parameters due to its non-negativity and its long tail. The mean value and variance of the above Gamma distribution are $a/b$ and $a/b^2$, respectively, which probabilistically control the magnitude of our hyper parameters $\tau$, $\{\lambda_r \}$ and $\{\gamma_{i_1 \ldots i_N}\}$. These hyper parameters then control $\{\mat{A^{(i)}}\}$ and $\ten{S}$. For instance, the noise term tends to have a very small magnitude if $\tau$ has a large mean value and a small variance; if $\lambda_r$ has a large mean value, then the $r$-th rank-1 term in the CP factorization tends to vanish, leading to rank reduction.

\subsection{Posterior Distribution of Model Parameters}
\begin{figure}[t]
\begin{center}
\begin{tikzpicture}[scale=0.6, every node/.style={scale=0.65},align = flush center,
data/.style={circle, draw, minimum size=1.6cm}
]
\matrix [
    matrix of nodes,
    column sep = 0.5em,
    row sep = 1.3em, 
    draw, dashed,
    nodes = {solid},
    ] (m)
{  
|[empty]|$a_0^\gamma$ &  |[empty]| & |[empty]|$b_0^\gamma$ & |[empty]|$c_0^\lambda$ & |[empty]| & |[empty]|$d_0^\lambda$ & |[empty]| & |[empty]|& |[empty]| \\
|[empty]| &  |[data]|$\gamma$ & |[empty]| & |[empty]| & |[data]|$\lambda$ & |[empty]| & |[empty]|$a_0^\tau$ & |[empty]|& |[empty]|$b_0^\tau$ \\
|[empty]| &  |[data]|$\mathcal{S}$ & |[empty]| & |[data]| $\bf{A}^{(1)}$ & |[empty]|$\dots$ & |[data]|$\bf{A}^{(N+1)}$ & |[empty]| & |[data]|$\tau$& |[empty]|\\
[1.6em]
|[empty]| &  |[empty]| & |[empty]| & |[empty]| & |[data]|$\mathcal{Y}_{\Omega_T}$ & |[data]|$\{\tilde{\mathcal{D}}_{t,\Omega_t}\}$ & |[empty]| & |[empty]|& |[empty]|\\[.7em]
};
\draw[->](m-1-1) to (m-2-2);
\draw[->](m-1-3) to (m-2-2);
\draw[->](m-2-2) to (m-3-2);
\draw[->](m-3-2) to (m-4-5);

\draw[->](m-1-4) to (m-2-5);
\draw[->](m-1-6) to (m-2-5);
\draw[->](m-2-5) to (m-3-4);
%\draw[->](m-2-5) to (m-3-5);
\draw[->](m-2-5) to (m-3-6);
\draw[->](m-3-4) to (m-4-5);
\draw[->](m-3-4) to (m-4-6);
\draw[->](m-3-6) to (m-4-5);
\draw[->](m-3-6) to (m-4-6);

\draw[->](m-2-9) to (m-3-8);
\draw[->](m-2-7) to (m-3-8);
\draw[->](m-3-8) to (m-4-6);
\draw[->](m-3-8) to (m-4-5);
\end{tikzpicture}
\caption{The probabilistic graphical model for our Bayesian robust streaming tensor completion.}
\label{fig:graphical model}
\end{center}
\end{figure}
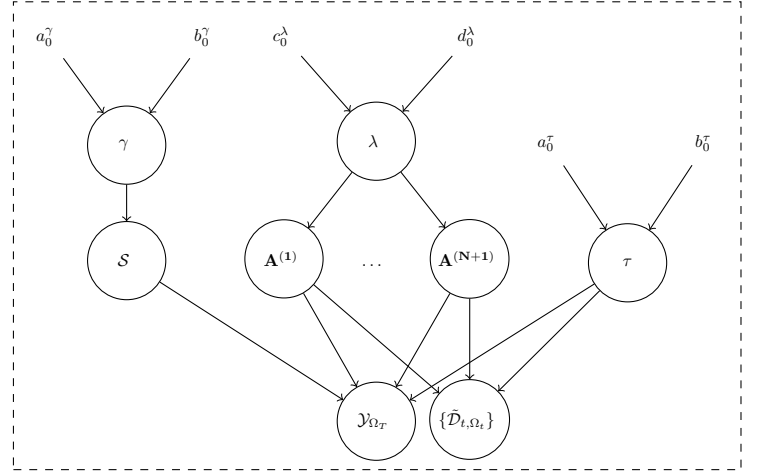

Now we can present a graphical model describing our Bayesian formulation in Fig.~\ref{fig:graphical model}. Our goal is to infer all hidden parameters based on partially observed data. 
For convenience, we denote all unknown hidden parameters in a compact form:
\begin{equation}
\Theta=\left \{ \{\mat{A}^{(n)}\}_{n=1}^{N+1},\ten{S}_{\Omega_T},\tau,\boldsymbol\lambda,\boldsymbol\gamma \right\}. \nonumber
\end{equation}
With the above likelihood function \eqref{eq:observationModel}, prior distribution for low-rank factors and sparse components in \eqref{eq:priorA} and \eqref{eq:priorS}, and prior distribution of the hyper-parameters in \eqref{eq:prior_hyper}, we can obtain the formulation of the posterior distribution in 
(\ref{eq:posterior density}) using Bayes theorem.
% \small
% \begin{multline}\label{eq:posterior density}
% p\left(\{\mat{A}^{(n)}\}_{n=1}^{N+1},\ten{S}_{\Omega_T},\tau,\boldsymbol\lambda,\gamma\middle\vert\ten{Y}_{\Omega_T},\{\ten{X}_{t,\Omega_t}\}_{t=i}^{T-1}\right) =\\
% p\left(\ten{Y}_{\Omega_T},\{\ten{X}_{t,\Omega_t}\}_{t=i}^{T-1}\middle\vert\{\mat{A}^{(n)}\}_{n=1}^{N+1},\ten{S}_{\Omega_T},\tau,\boldsymbol\lambda,\gamma \right)p\big(\mat{A}^{(n)}\big\vert \boldsymbol\lambda \big)p(\boldsymbol\lambda)
% p(\ten{S}_{\Omega_T}\vert\boldsymbol\gamma)p(\boldsymbol\gamma)
% \end{multline}
% \normalsize

% \begin{strip}
% \begin{equation}\label{eq:posterior density}
% p\left(\Theta\middle\vert\ten{Y}_{\Omega_T},\{\tilde{\ten{X}}_{t,\Omega_t}\}\right) =
% \frac{p\left(\ten{Y}_{\Omega_T},\{\tilde{\ten{X}}_{t,\Omega_t}\}\middle\vert\ \Theta\right){\prod_{n=1}^{(N+1)}p\big(\mat{A}^{(n)}}\big\vert \boldsymbol\lambda \big)p(\boldsymbol\lambda)
% p(\ten{S}_{\Omega_T}\vert\boldsymbol\gamma)p(\boldsymbol\gamma)p(\tau)}{p(\ten{Y}_{\Omega_T},\{\tilde{\ten{X}}_{t,\Omega_t}\})}.
% \end{equation}
% \end{strip}

The main challenge is how to estimate the resulting posterior distribution (\ref{eq:posterior density}). We address this issue in Section~\ref{sec:update}.

\section{Variational Bayesian Solver For Model Parameter Estimation}
\label{sec:update}
It is hard to obtain the exact posterior distribution (\ref{eq:posterior density}) because the marginal density $p(\ten{Y}_{\Omega_T},\{\tilde{\ten{D}}_{\Omega_t}\})$ is unknown and is expensive to compute. Therefore, we employ variational Bayesian inference~\cite{winn2005variational} to obtain a closed-form approximation of the posterior density (\ref{eq:posterior density}). The variational Bayesian method was previously employed for matrix completion~\cite{babacan2012sparse} and non-streaming tensor completion~\cite{zhao2015bayesian, zhao2016bayesian}, and it is a popular inference technique in many domains. We use a similar procedure to ~\cite{babacan2012sparse, zhao2015bayesian} to derive our iteration steps, but the details are quite different since we solve a streaming problem and we approximate an entirely different posterior distribution. 

Due to the complexity of the updates, we defer these derivations to Section \ref{sec: updates}. In this section, we only provide some key results and intuitions.

\subsection{Variational Bayesian}

Our goal is to find a distribution $q(\Theta)$ that approximates the true posterior distribution $p(\Theta\vert \ten{Y}_{\Omega_T},\{\tilde{\ten{D}}_{\Omega_t}\})$ by minimizing the KL divergence. The KL divergence between two distributions is defined by
\begin{equation}\label{eq:VBKL}
\begin{split}
\small
\text{KL}\big(q(\Theta)\big\vert\big\vert p(\Theta\vert \ten{Y}_{\Omega_T},\{\tilde{\ten{D}}_{\Omega_t}\})\big)= \ln p(\ten{Y}_{\Omega_T},\{\tilde{\ten{D}}_{\Omega_t}\}) &- \mathcal{L}(q), \\
 \mbox{where} \; \mathcal{L}(q) = \int q(\Theta)\ln \left(\frac{p(\ten{Y}_{\Omega_T}, \{\tilde{\ten{D}}_{\Omega_t}\},\Theta)}{q(\Theta)}\right)&d\Theta.
\end{split}
\end{equation}

The quantity $\ln p(\ten{Y}_{\Omega_T},\{\tilde{\ten{D}}_{\Omega_t}\})$ denotes model evidence and is a constant. Therefore, minimizing the KL divergence is equivalent to maximizing $\mathcal{L}(q)$. To do so we apply the mean field variational approximation ~\cite{bishop2014pattern}. That is, we assume that the posterior can be factorized as a product of the individual marginal distributions:
\begin{equation}
\label{eq:vbfactorization}
q\left(\Theta\right) =\left\{ \prod_{n=1}^{N+1} q\left(\mat{A}^{(n)}\right) \right\}q(\ten{S}_{\Omega_T}) q(\boldsymbol\lambda)q(\boldsymbol\gamma) q(\tau).
\end{equation}
where $\Theta$ is the collection of all parameters. The main advantage of this assumption is that we can maximize $\mathcal{L}(q)$, and therefore optimize KL divergence, by applying an alternating update rule to each factor in turn. The update rule for an individual parameter $\Theta_i$ is given by
\begin{equation}\label{eq: param update}
\ln q(\Theta_i) \propto \mathbb{E}_{\Theta_{j\neq i}}\ln (p(\ten{Y}_{\Omega_T},\{\tilde{\ten{D}}_{\Omega_t}\},\Theta)),
\end{equation}
where the subscript $\Theta_{j\neq i}$ denotes the expectation with respect to all latent factors except $\Theta_i$.

In the following we will provide the closed-form expressions of these alternating updates. 

\subsection{Factor Matrix Updates}

The posterior distribution of an individual factor matrix is
\begin{equation*}\label{eq:psoteriorA}
q\big(\mat{A}^{(n)}) = \prod_{i_n=1}^{I_n} \mathcal{N}\big(\hat{\mat{a}}^{(n)}_{i_n}\big\vert \bar{\mat{a}}_{i_n}^{(n)},\mat{V}_{i_n}^{(n)}\big).\end{equation*}
Note that $\hat{\mat{a}}^{(n)}_{i_n}$ denotes the $i_n$th row of $\mat{A}^{(n)}$. Therefore, we only need to update the posterior mean $\bar{\mat{a}}_{i_n}^{(n)} \in \mathbb{R}^{R}$ and co-variance matrix $\mat{V}_{i_n}^{(n)} \in \mathbb{R}^{R\times R}$.

{\bf Update non-temporal factors.} {All non-time factors are updated by Equations (\ref{eq:updatevar}) and (\ref{eq:update non time matrix}). Notationally, this means that the value $n$ ranges in the set $\{1,\dots,N\}$ for the two updates below.

\begin{equation}\label{eq:updatevar}
\small
{\mat{V}}^{(n)}_{i_n} = \left(\mathbb{E}_q[\tau] \sum_{t=i}^T{\mu^{T-i}}\mathbb{E}_q\left[\mat{A}_{i_n}^{(\setminus n)T}\mat{A}_{i_n}^{(\setminus n)}\right]_{\Omega_t} +\mathbb{E}_q[\boldsymbol\Lambda]\right)^{-1},
\end{equation}
\begin{multline}\label{eq:update non time matrix}
\small
{\bar{\mat{a}}}^{(n)}_{i_n} = \mathbb{E}_q[\tau]\mat{V}^{(n)}_{i_n}\Bigg(\mathbb{E}_q\left[\mat{A}_{i_n}^{(\setminus n)T}\right]_{\Omega_T}\text{vec}\left( \ten{Y}_{\Omega_T}-\mathbb{E}_q[\ten{S}_{\Omega_T}]\right) \\
+\sum_{t=i}^{T-1} \mu^{T-t}\mathbb{E}_q\left[\mat{A}_{i_n}^{(\setminus n)T}\right]_{\Omega_t}\text{vec}\left(\tilde{\ten{D}}_{\Omega_t,i_n}\right) \Bigg).
\end{multline}
The double subscript $\{\Omega_t, i_n\}$ represents the sampled mode $n-1$ subtensor obtained by fixing index $n$ to $i_n$. The notation $\mathbb{E}_q\left[\mat{A}_{i_n}^{(\setminus n)}\right]_{\Omega_t}$ represents a sampled expectation of the excluded Khatri-Rao product:
\[
\mathbb{E}_q\left[\mat{A}_{i_n}^{(\setminus n)}\right]_{\Omega_t} = \left(\mathbb{E}_q\left[\bigodot_{j\neq n} \mat{A}^{(j)}\right]\right)_{\mathbb{I}_{i_n}}.
\]
The matrix $\mat{A}_{i_n}^{(\setminus n)}$ is $\prod_{j\neq n}I_j\times R$ and the indicator function ${\mathbb{I}_{i_n}}$ samples the row $(i_1,\dots, i_{n-1},i_{n+1},\dots, i_{N+1})$ if the entry $(i_1,\dots, i_{n-1},i_n,i_{n+1},\dots,i_{N+1})$ is in $\Omega_t$ and sets the row to zero if not. The expression $\mathbb{E}_q[\cdot]$ denotes the posterior expectation with respect to all variables involved. 

{\bf Update temporal factors.} The temporal factors require a different update scheme because the factors corresponding to different time slices do not interact with each other. For all time factors the variance is updated according to
\begin{align}\label{eq:updatetimevariance}
\small
\begin{split}
{\mat{V}}^{(N+1)}_{t-i+1} = \left(\mathbb{E}_q[\tau] {\mu^{T-t}}\mathbb{E}_q\left[\mat{A}_{t-i+1}^{(\setminus (N+1))T}\mat{A}_{t-i+1}^{(\setminus (N+1))}\right]_{\Omega_t} +\mathbb{E}_q[\mat{\Lambda}]\right)^{-1}.
\end{split}
\end{align}
The rows of the time factor matrix are updated differently depending on the slice in question. Since we assume that past observations have had their sparse errors removed, the time factors of all past slices (so $t\neq T$) can be updated by
\begin{align}\label{eq:update time matrix}
\small
\begin{split}
{\bar{\mat{a}}}^{(N+1)}_{t-i+1} = \mathbb{E}_q[\tau]\mat{V}^{(N+1)}_{t-i+1}\left(\mu^{T-t}\mathbb{E}_q\left[\mat{A}_{t-i+1}^{(\setminus N+1)T}\right]_{\Omega_t}\text{vec}\left(\tilde{\ten{D}}_{\Omega_t}\right) \right).
\end{split}
\end{align}
The factors corresponding to time slice $T$ depend on the sparse errors removed in the current step. The update is therefore given by
\begin{align}\label{eq:updatetimevariance2}
\small
\begin{split}
{\bar{\mat{a}}}^{(N+1)}_{T-i+1} = \mathbb{E}_q[\tau]\mat{V}^{(N+1)}_{T-i+1}\left(\mathbb{E}_q\left[\mat{A}_{T-i+1}^{(\setminus N+1)T}\right]_{\Omega_T}\text{vec}\left(\ten{Y}_{\Omega_T}-\mathbb{E}_q\left[\ten{S}_{\Omega_T}\right]\right) \right).
\end{split}
\end{align}

{\bf Intuition.} The update terms are rather dense so we provide some intuitions. We update the variance $\mathbf{V}^{(n)}_{i_n}$ by combining $\mathbb{E}_q[\boldsymbol\Lambda]$, denoting the factor prior, and covariance of other factor matrices. The tradeoff between these two terms is controlled by $\mathbb{E}_q[\tau]$, which denotes precision, or the current fitness of the model. If the current model fitness is high then the information received from the prior is weighted less heavily. The $\bar{\mat{a}}^{(n)}_{i_n}$ update is formed by finding a row vector that maximizes model fit across all elements of the sliding window. The outcome is then rescaled by the model fitness and rotated by the covariance $\mathbf{V}^{(n)}_{i_n}$.

Evaluating the expectation of the Khatri-Rao product in the preceding updates is challenging. This computation is addressed in Lemma IV.3 of \cite{zhao2016bayesian}. We provide the result below.
\begin{equation}
\label{eq:ENATA}
\mathbb{E}_q\big[\mat{A}_{i_n}^{(\backslash n)T}\mat{A}_{i_n}^{(\backslash n)}\big]_{\Omega_t} = \!\!\!\!\!\!\!\!\sum_{(i_1,\ldots,i_N)\in\Omega_t} \hadamard_{k\neq n} \left(\mathbb{E}_q\left[ \hat{\mat{a}}^{(k)}_{i_k}\hat{\mat{a}}^{(k)T}_{i_k}\right] \right).
\end{equation}

The row-wise expectation can be evaluated as follows: let $\mathbf{B}^{(n)}$ of size $I_n\times R^2$ denote an expectation of a quadratic form related to $\mat{A}^{(n)}$ by defining $i_n$th-row vector
\begin{equation}\label{eq: row expectation}
\mathbf{b}^{(n)}_{i_n} = \text{vec}\left(\mathbb{E}_q\left[ \hat{\mat{a}}^{(n)}_{i_n}\hat{\mat{a}}^{(n)T}_{i_n}\right]\right) = \text{vec}\left( {\bar{\mat{a}}}^{(n)}_{i_n}{\bar{\mat{a}}}^{(n)T}_{i_n} + \mathbf{V}^{(n)}_{i_n}\right).
\end{equation}

Then (\ref{eq:ENATA}) can be written as
\begin{equation*}\label{eq:ATA}
\text{vec}\left(\mathbb{E}_q\big[\mat{A}_{i_n}^{(\backslash n)T}\mat{A}_{i_n}^{(\backslash n)}\big]_{\Omega_t}\right)
= \Big(\bigodot_{k\neq n}\mathbf{B}^{(k)}\Big)^T \; \text{vec}(\ten{O}_{t}).
\end{equation*}
where the tensor $\ten{O}_t$ is an indicator tensor constructed from the sampled entries $\Omega_t$. 

\subsection{Posterior Distribution of Hyperparameters $\boldsymbol\lambda$}

The posteriors of the parameters $\lambda_r$ are independent Gamma distributions. Therefore the joint distribution takes the form
\[
q(\boldsymbol\lambda) = \prod_{r=1}^{R} \text{Ga}(\lambda_r \vert {c}_M^r, {d}_M^r )
\]
where $c_M^r$, $d_M^r$ denote the posterior parameters learned from the previous $M$ iterations. The updates to $\boldsymbol{\lambda}$ are given below.
\begin{equation}\label{eq:update lambda}
c_M^r = c_0+1 + \frac{1}{2}\sum_{n=1}^{N} I_n, \quad
d_M^r = d_0 + \frac{1}{2}\sum_{n=1}^{N+1} \mathbb{E}_q\left[{\mat{a}}^{(n)T}_{ r}{\mat{a}}^{(n)}_{ r}\right]
\end{equation}
We note that the vectors $\mat{a}^{(n)T}_{ r}$ are the columns of the factor matrix $\mat{A}^{(n)}$ rather than the row vectors we used in prior computations. The updates given in Equation (\ref{eq:update lambda}) enforce sparsity as follows: large columns corresponding to factor $r$ increase the rate parameter $d^r_M$. This decreases $\lambda_r$. Then the inversion in Equation (\ref{eq:updatevar}) that creates the variance matrix assigns the $r^{th}$ low-rank factor a high variance, and therefore a higher probability of being nonzero. 

The expectation term in (\ref{eq:update lambda}) can be evaluated using a similar computation to (\ref{eq: row expectation}).
\[\mathbb{E}_q\left[{\mat{a}}^{(n)T}_{r}{\mat{a}}^{(n)}_{ r}\right] = \mathbb{E}_q\left[{{\mat{a}}}^{(n)T}_{r} \right]\mathbb{E}_q\left[{{\mat{a}}}^{(n)}_{r}\right] + \sum_{i_n} \left(\mathbf{V}_{i_n}^{(n)}\right)_{rr}
\]
Then the second update in Equation (\ref{eq:update lambda}) can be written in matrix form by updating $\mathbf{d}_M=[d_M^1,\ldots d_M^R]^T$ with
\begin{equation*}\label{eq:lambda3}
\mathbf{d}^{(n)}_M = d_0+\frac{1}{2}\left(\text{diag}\left({\bar{\mat{A}}}^{(n)T}{\bar{\mat{A}}}^{(n)} + \sum_{i_n} \mathbf{V}_{i_n}^{(n)} \right) \right).
\end{equation*}
The notation ${\bar{\mat{A}}}^{(n)}$ denotes the posterior mean of the entire factor matrix. The expectation of each rank-sparsity parameter can then be computed as
\[
\mathbb{E}_q[\boldsymbol\Lambda] = \text{diag}([c_M^1/d_M^1,\ldots,c_M^R/d_M^R]).
\]

\subsection{Posterior Distribution of Sparse tensor $\ten{S}$}
The posterior approximation of $\ten{S}_{\Omega_T}$ is given by
\begin{equation}
q(\ten{S}_{\Omega_T}) = \prod_{(i_1,\ldots,i_N)\in\Omega_T} \normalpdf{\mathcal{S}_{i_1 \ldots i_N} }{\bar{\mathcal{S}}_{i_1 \ldots i_N}}{\sigma^2_{i_1 \ldots i_N}},
\end{equation}
where the posterior parameters can be updated by
\begin{gather}\label{eq:update S}
\begin{aligned}
{\bar{\mathcal{S}}_{i_1 \ldots i_N}} &= \sigma^2_{i_1 \ldots i_N}\mathbb{E}_q[\tau]\Big(\mathcal{Y}_{i_1\ldots i_N}-\\
&\mathbb{E}_q\left[\left\langle\hat{\mat{a}}_{i_1}^{(1)},\ldots,\hat{\mat{a}}_{i_N}^{(N)};\hat{\mat{a}}_{T-i+1}^{(N+1)}\right\rangle \right] \Big) \\
\sigma^2_{i_1 \ldots i_N} &= (\mathbb{E}_q[\gamma_{i_1\ldots i_N}] + \mathbb{E}_q[\tau])^{-1}.
\end{aligned}
\end{gather}

The sparse tensor $\ten{S}_{\Omega_T}$ picks out entries that are not well-described by the expectation of the CP factors. The size of sparse entries is governed by the prior expectation $\mathbb{E}_q[\gamma_{i_1\ldots i_N}]$ and the determined precision of Gaussian noise $\mathbb{E}_q[\tau]$. The sparse term represents a tradeoff governed by the noise precision prior $\tau$ and rank-sparsity parameter $\boldsymbol\lambda$. The CP factors explain as much of the data as as possible given $\boldsymbol\lambda$ and the unexplained data is absorbed into the sparse error term $\ten{S}_{\Omega_T}$.

\subsection{Posterior Distribution of Hyperparameters $\boldsymbol\gamma$}

The posterior of $\boldsymbol\gamma$ is also factorized into entry-wise independent distributions
\begin{equation}
q(\boldsymbol\gamma) = \prod_{(i_1,\ldots,i_N)\in\Omega_T}\text{Ga}(\gamma_{i_1 \ldots i_N} \vert {a}_M^{\gamma_{i_1 \ldots i_N}}, {b}_M^{\gamma_{i_1 \ldots i_N}} ),
\end{equation}
whose posterior parameters can be updated by
\begin{equation}\label{eq:update gamma}
{a}_M^{\gamma_{i_1 \ldots i_N}} = a_0^\gamma + \frac{1}{2}, \hspace{.04in}
{b}_M^{\gamma_{i_1 \ldots i_N}} = b_0^\gamma + \frac{1}{2}({\bar{\mathcal{S}}}^2_{i_1 \ldots i_N} + \sigma^2_{i_1 \ldots i_N}).
\end{equation}
Smaller values of $\bar{\mathcal{S}}^2_{i_1 \ldots i_N}$ enforce larger values $\mathbb{E}_q[\gamma_{i_1 \ldots i_N}]$ which enforce ${\mathcal{S}}_{i_1 \ldots i_N}$ to be zero by (\ref{eq:update S}), and vice versa. Therefore large elements of $\ten{S}_{\Omega_T}$ posses more inertia while smaller elements are forced towards zero. Sparsity of  $\ten{S}_{\Omega_T}$ must be strongly enforced to prevent the sparse error term from explaining the entirety of the data via a series of entrywise independent Gaussians.

\subsection{Posterior Distribution of Parameter $\tau$}
The posterior PDF of the noise precision is again a Gamma distribution. The noise precision is controlled by the model residuals, and the posterior parameters can be updated by
\begin{gather}\label{eq:update tau}
\begin{aligned}
\small
&{a}_M^\tau = a_0^\tau+\frac{1}{2}\sum_{t=i}^T|\Omega_t| ,\\
&{b}_M^\tau = b_0^\tau+  \frac{1}{2}\mathbb{E}_q\left[\left\|\left(\ten{Y}-{\ten{S}}- [\![ {\mat{A}}^{(1)},\ldots,{\mat{A}}^{(N)}; \hat{{\mat{a}}}_{T-i+1}^{(N+1)} ]\!] \right)_{\Omega_T}\right\|_{\rm F}^2\right]\\
&+\frac{1}{2}\mathbb{E}_q\left[\sum_{t=i}^{T-1} \mu^{T-t}\left\|\left(\tilde{\ten{D}}_t-[\![ {\mat{A}}^{(1)},\ldots,{\mat{A}}^{(N)}; \hat{{\mat{a}}}_{t-i+1}^{(N+1)} ]\!] \right)_{\Omega_t} \right\|_F^2\right].
\end{aligned}
\end{gather}

The Frobenius norm terms control the noise precision $\tau$  through the rate parameter $b_M^\tau$. An increase in $b_M^\tau$ occurs when the model does not explain the data well. This results in a decrease in the precision since $
\mathbb{E}[\tau] = \frac{a_M^\tau}{b_M^\tau}$.
The shape parameter $a^\tau_M$ weights the residuals by the number of considered entries. The update of the noise term is the most expensive update as the size of the tensor grows. In order to avoid excessive computation we update $\tau$ based on only the current slice. We view this as a noisy estimate of the true update, which is a weighted sum across several previous slices. The expectation of the residuals in Equation (\ref{eq:update tau}) is challenging to compute so we present several results from \cite{zhao2016bayesian}.

\begin{lemma}\label{theorem:2}
Given a set of independent random matrices $\{\mat{A}^{(n)}|n=1,\ldots,N\}$, we assume that $\forall n, \forall i_n$, the row vectors $\{\mat{a}^{(n)}_{i_n}\}$ are independent, then
\begin{multline*}\label{eq:expX}
\mathbb{E}\left[ \left\| [\![ \mat{A}^{(1)},\ldots, \mat{A}^{(N)};\hat{\mat{a}}_{T-i+1}^{(N+1)} ]\!]\right\|_F^2 \right]\\ = \sum_{i_1,\ldots,i_N} \bigg\langle \mathbb{E}\left[\hat{\mat{a}}_{i_1}^{(1)}\hat{\mat{a}}_{i_1}^{(1)T}\right],\ldots
,\mathbb{E}\left[\hat{\mat{a}}_{i_N}^{(N)}\hat{\mat{a}}_{i_N}^{(N)T}\right],\\
\mathbb{E}\left[\hat{\mat{a}}_{T-i+1}^{(N+1)}\hat{\mat{a}}_{T-i+1}^{(N+1)T}\right]\bigg\rangle.
\end{multline*}
\end{lemma}
Lemma \ref{theorem:2} allows for evaluation of the current slice residual error term from Equation (\ref{eq:update tau}) via
\begin{equation*}\label{eq:modelerror}
\begin{split}
&\mathbb{E}_q\left[\left\|\left(\ten{Y}-[\![ \mat{A}^{(1)},\ldots, \mat{A}^{(N)};\hat{\mat{a}}^{(N+1)}_{T-i+1}]\!]-\ten{S}\right)_{\Omega_T}\right\|_F^2\right]\\
=&\|\ten{Y}_{\Omega_T}\|_F^2 - 2\text{vec}^T(\ten{Y}_{\Omega_T})\text{vec}\left([ \![ \bar{\mat{A}}^{(1)},\ldots,\bar{\mat{A}}^{(N)};\bar{\mat{a}}^{(N+1)}_{T-i+1} ]\!]_{\Omega_T} \right) \\
&+ \text{vec}^T(\ten{O}_T) \left(\bigodot_n \mathbf{B}^{(n)} \right)\mathbf{1}_{R^2} - 2\text{vec}^T(\ten{Y}_{\Omega_T})\text{vec}(\bar{\ten{S}}_{\Omega_T}) \\
& +2\text{vec}^T([\![\bar{\mat{A}}^{(1)},\ldots,\bar{\mat{A}}^{(N)};\bar{\mat{a}}^{(N+1)}_{T-i+1} ]\!]_{\Omega_T})\text{vec}(\bar{\ten{S}}_{\Omega_T}) \\
& +\mathbb{E}_q[\|\ten{S}_{\Omega_T}\|_F^2].
\end{split}
\end{equation*}
where $\mathbf{1}_{R^2}$ is a length $R^2$ column vector of ones.

% \subsubsection{Lower bound of model evidence}
% One can also estimate the variational lower bound from Equation (\ref{eq:VBKL}) in order to update the hyperparameters controlling the noise. The lower bound is
% \begin{equation}\label{eq:lowerbound}
% \mathcal{L}(q) = \mathbb{E}_{q(\Theta)}[\ln p(\ten{Y}_{\Omega_T},\Theta)] + H(q(\Theta)),
% \end{equation}
% where the first term denotes the posterior expectation of joint probability density, and the second term denotes the entropy of $q$ distribution.

% We can then update the hyperparameters controlling the noise by optimizing
% \begin{multline}\label{eq:updatetoplevel}
% \mathcal{L}(a_0^\gamma,b_0^\gamma) = -M\ln\Gamma(a_0^\gamma) + Ma_0^\gamma \ln b_0^\gamma \\ +(a_0^\gamma-1)\sum_{(i_1\ldots i_N)\in\Omega} \left\{(\psi(a_M^{\gamma_{i_1\ldots i_N}}) -\ln b_M^{\gamma_{i_1\ldots i_N}})\right\}\\
% -b_0^\gamma \sum_{(i_1\ldots i_N)\in\Omega}\
% \frac{a_M^{\gamma_{i_1\ldots i_N}}}{b_M^{\gamma_{i_1\ldots i_N}}}.
% \end{multline}

\subsection{Algorithm}
We provide the algorithmic details for our model. The same algorithm applies for the factorization of complete or incomplete data. In the case of a complete tensor, each $\Omega_t$ contains all possible indices.
\begin{figure*}[t]
\centering     %%% not \center
\subfigure[Relative errors of factorizing the full tensor.]{\label{fig:synthetic factorization}\includegraphics[width=.47\textwidth]{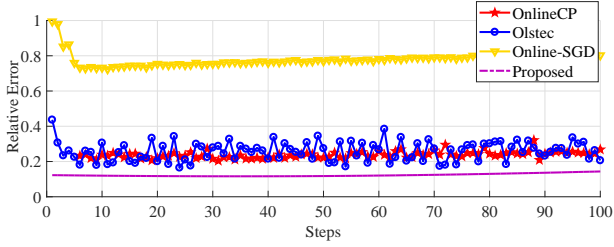}} \hspace{0.1in}
\subfigure[Relative errors of tensor completion based on $15\%$ samples.]{\label{fig:synthetic completion}\includegraphics[width=.47\textwidth]{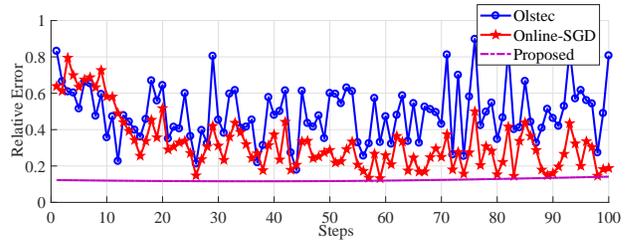}}
\caption{Comparison to existing streaming tensor factorization and completion algorithms on synthetic data with 100 rank-5 $40\times 40$ slices and with $1\%$ of entries corrupted.}
\label{fig: big synthetic}
\end{figure*}

\paragraph{Initialization}
Since variational Bayesian inference is only guaranteed to converge to a local minimum, a good initialization is important. We follow the initialization of \cite{zhao2016bayesian}. The hyperparameters are initialized by $\mathbb{E}[\boldsymbol\Lambda]=\mathbf{I}$, $\mathbb{E}[\tau]=1$ and $\forall n, \forall i_n,\mathbb{E}[\boldsymbol\gamma_{i_1\ldots i_N}] = 1$. For the factor matrices, $\mathbb{E}[\mat{A}^{(n)}]$ is set to $\mat{A}^{(n)}= \mathbf{U}^{(n)}\boldsymbol\Sigma^{(n)^{\frac{1}{2}}}$, where $\mathbf{U}^{(n)}$ denotes the left singular vectors and $\boldsymbol\Sigma^{(n)}$ denotes the diagonal singular values matrix, obtained by SVD of \mbox{mode-$n$} matricization of $\ten{Y}$. $\mathbf{V}^{(n)}$ is set to $\mathbb{E}[\boldsymbol\Lambda^{-1}]$. For the sparse tensor $\ten{S}$, $\mathbb{E}[\mathcal{S}_{i_1\ldots i_N}]$ is drawn from $\mathcal{N}(0,1)$, while $\sigma^2_{i_1\ldots i_N}$ is set to $\mathbb{E}[\boldsymbol\gamma^{-1}_{i_1\ldots i_N}]$. The tensor rank $R$ is initialized by the maximum rank $R\leq \min_n P_n$, where $P_n= \prod_{i\neq n} I_i$. In practice one manually sets a maximum allowable rank via the initialization value of $R$. The final rank discovered does not depend on the initialization value, as long as the initialization value is high enough.
\paragraph{Iterative Process}
The overall flow of our algorithm amounts to collecting the individual update terms in sequence. We stop iterating and declare our update scheme converged when the change in the variational lower bound from Equation (\ref{eq:VBKL}) is less then $10^{-4}$ per iteration. Our algorithm is summarized in Algorithm \ref{algoflow}.
\begin{algorithm}
    \caption{Variational Bayesian Updating Scheme for Streaming Tensor Completion}
    \label{algoflow}
    \begin{algorithmic}
    \WHILE{Not Converged}
\STATE Update the variance matrices via Equations (\ref{eq:updatevar},\ref{eq:updatetimevariance})
\STATE Update the factor matrices by Equations (\ref{eq:update non time matrix},\ref{eq:update time matrix}, \ref{eq:updatetimevariance2})
\STATE Update the rank prior $\boldsymbol \lambda$ by Equation (\ref{eq:update lambda})
\STATE Update the sparse term $\ten{S}_{\Omega_T}$ by Equation (\ref{eq:update S})
\STATE Update the sparsity prior $\boldsymbol  \gamma$ by Equation (\ref{eq:update gamma})
\STATE Update the precision $\tau$ by Equation (\ref{eq:update tau})
\ENDWHILE

    \end{algorithmic}
\end{algorithm}

% \begin{figure}\label{fig: synthetic factorization}
% \centering{
% \includegraphics[width = 2in]{figures/synthetic_comparison}
% }
% \caption{Relative Reconstruction Error}
% \end{figure}
\section{Derivations of the Update Process}\label{sec: updates}

In this section we provide the main steps of deriving our factor matrix updates and the noise term update. The other updates can be derived from results in the appendix of~\cite{zhao2016bayesian}. In order to reduce the complexity of the factor matrix update calculations we introduce several new pieces of notation. The new notation will allow us to extract a single factor matrix row $\hat{\mat{a}}_{i_n}^{(n)}$ from complicated expressions. 

We represent the low-rank estimate at time $t \in [i, T]$ by
\[
\ten{A}_t  = [\![ \mat{A}^{(1)},\dots, \mat{A}^{(N)},\hat{\mat{a}}^{(N+1)}_{t-i+1}]\!].
\]
We also introduce a time index to the excluded Khatri-Rao product:
\[
\bigodot_{\substack{k\neq n \\ t-i+1}} \mat{A}^{(k)}  = \left( \bigodot_{k\neq n} \mat{A}^{(k)} \right)\bigotimes \hat{\mat{a}}^{(N+1)}_{t-i+1}.
\]
Next, we introduce a notation for the sampled inner product of two tensors:
\begin{equation}\label{eq: tensor inner product}
\langle\ten{B}, \ten{A}_t\rangle_{\Omega_t} = \text{vec}(\ten{B}_{\Omega_t})^T\text{vec}(\ten{A}_{\Omega_t})
\end{equation}
This notation will allow us to express the squared sampled Frobenius norm $\|(\ten{B}-\ten{A}_t)_{\Omega_t}\|^2_F$ in a compact format. For our purposes $\ten{B}$ will be a constant data tensor, i.e. $\ten{\tilde{D}}_t$. We will use the subscript $i_n$, for example $\ten{A}_{i_n}$, to denote the order $N-1$ subtensor of $\ten{A}$ obtained by restricting the $n^{th}$ index to $i_n$. 

We now collect the preceding notations and present a lemma. This lemma will be used when we derive the update process later in this section. 

\begin{lemma}\label{lem: squared frobenius} Let $\mathbb{E}_q$ denote the expectation with respect to all variables except $\hat{\mat{a}}^{(n)}_{i_n}$. Then we can compute two expectations:
\begin{multline}\label{eq: lemma part 1}
\mathbb{E}_q\left[\|\ten{A}_{\Omega_t} \|_F^2\right] =\\ \hat{\mat{a}}^{(n)}_{i_n}\mathbb{E}_q\left[\left( \bigodot_{\substack{k\neq n \\ t-i+1}} \mat{A}^{(k)}\right)_{\Omega_t}^T \left( \bigodot_{\substack{k\neq n \\ t-i+1}} \mat{A}^{(k)}\right)_{\Omega_t}\right] \hat{\mat{a}}^{(n)T}_{i_n}+\text{const}.
\end{multline}
\begin{multline}\label{eq: lemma part 2}
\mathbb{E}_q\left[\langle\ten{B}, \ten{A}_t\rangle_{\Omega_t}\right] = \hat{\mat{a}}^{(n)}_{i_n}\left( \bigodot_{\substack{k\neq n \\ t-i+1}} \bar{\mat{A}}^{(k)}\right)_{\Omega_t}^T \textnormal{vec}(\ten{B}_{\Omega_t,i_n})+\text{const}
\end{multline}
\end{lemma}
The constant term denotes all quantities constant with respect to $\hat{\mat{a}}_{i_n}^{(n)}$. The expectation in (\ref{eq: lemma part 1}) can be computed using Equation (\ref{eq:ENATA}).

\begin{proof} For both computations we split the tensor $\ten{A}_t$ into subtensors. The vectorized order $N-1$ subtensor obtained from $\ten{A}_{\Omega_t}$ by fixing index $n$ to $i_n$ is given by
\begin{align}\label{}
\text{vec}\left(\ten{A}_{\Omega_t,i_{n}}\right)= \hat{\mat{a}}^{(n)}_{i_n}\left( \bigodot_{\substack{k\neq n \\ t-i+1}} \mat{A}^{(k)}\right)_{\Omega_t}^T 
\end{align}
This allows us to compute the squared Frobenius norm
\begin{align}\label{eq: appendixeq1}
\begin{split}
\|\ten{A}_{\Omega_t,i_{n}} \|_F^2 &= 
\text{vec}\left(\ten{A}_{\Omega_t,i_{n}}\right)^T \text{vec}\left(\ten{A}_{\Omega_t,i_{n}}\right) \\
&= \hat{\mat{a}}^{(n)}_{i_n}\left( \bigodot_{\substack{k\neq n \\ t-i+1}} \mat{A}^{(k)}\right)_{\Omega_t}^T \left( \bigodot_{\substack{k\neq n \\ t-i+1}} \mat{A}^{(k)}\right)_{\Omega_t} \hat{\mat{a}}^{(n)T}_{i_n}.
\end{split}
\end{align}
We note that 
\[
\|\ten{A}_{\Omega_t} \|_F^2 = \|\ten{A}_{\Omega_t,i_{n}} \|_F^2+\sum_{i_j \neq i_n} \|\ten{A}_{\Omega_t,i_{j}} \|_F^2.
\]
Of the terms on the right hand side, only $\|\ten{A}_{\Omega_t,i_{n}} \|_F^2$ depends on $\hat{\mat{a}}_{i_n}^{(n)}$. Then we can take the expectation as in Lemma \ref{lem: squared frobenius}.  
\begin{align}
\begin{split}
\mathbb{E}_q\left[\|\ten{A}_{\Omega_t} \|_F^2\right] &= \mathbb{E}_q\left[\|\ten{A}_{\Omega_t,i_{n}} \|_F^2\right]+\text{const}
\end{split}
\end{align}
Applying (\ref{eq: appendixeq1}) proves part one of Lemma \ref{lem: squared frobenius}.

To prove part two first we decompose $\ten{B}$ and $ \ten{A}_t$ into their subtensors: 
\begin{align*}
\begin{split}
&\langle\ten{B}, \ten{A}_t\rangle_{\Omega_t}= \sum_{i_j = 1}^{I_n} \langle\ten{B}_{i_j}, \ten{A}_{t,i_j}\rangle_{\Omega_t}\\
&=\hat{\mat{a}}^{(n)}_{i_n}\left( \bigodot_{\substack{k\neq n \\ t-i+1}} {\mat{A}}^{(k)}\right)_{\Omega_t}^T \textnormal{vec}(\ten{B}_{\Omega_t,i_j})+\sum_{i_j \neq i_n} \langle\ten{B}_{i_j}, \ten{A}_{t,i_j}\rangle_{\Omega_t}\\
&=\hat{\mat{a}}^{(n)}_{i_n}\left( \bigodot_{\substack{k\neq n \\ t-i+1}} {\mat{A}}^{(k)}\right)_{\Omega_t}^T \textnormal{vec}(\ten{B}_{\Omega_t,i_n})+ \sum_{i_j \neq i_n} \langle\ten{B}_{i_j}, \ten{A}_{t,i_j}\rangle_{\Omega_t}
\end{split}
\end{align*}
The 2nd-term on the right-hand side is independent with respect to $\hat{\mat{a}}_{i_n}^{(n)}$, so the expectation is a constant. This proves part two of Lemma \ref{lem: squared frobenius}.
\end{proof}

Based on Lemma \ref{lem: squared frobenius}, Equation (\ref{eq: big compute}) shows the detailed derivation for the update formulation of the non-temporal factor $\hat{\mat{a}}_{i_n}^{(n)}$ given in Equations (\ref{eq:updatevar}) and (\ref{eq:update non time matrix}). The variational posterior of $\hat{\mat{a}}^{(n)}_{i_n}$ is normal, therefore our goal is to extract the sufficient statistics of the Gaussian distribution. At each step of the computation we move all terms that are independent of $\hat{\mat{a}}^{(n)}_{i_n}$ into the constant term. In the first two lines we provide an expression for the log-likelihood, computed from our posterior distribution in Equation (\ref{eq:posterior density}). Then we expand the Frobenius norm terms so that we can apply Lemma (\ref{lem: squared frobenius}). Next we factor and regroup so that our expression takes the form of a Gaussian. Finally, we extract the sufficient statistics by the method of ``completing the square"~\cite{bishop2014pattern}. All other non-temporal factor updates can be derived in the same way. 
\begin{figure*}%[b]
\begin{align}\label{eq: big compute}
\small
\begin{split}
\ln q(\bf{\hat{a}_{i_n}^{(n)}})=&\mathbb{E}_{q(\Theta \setminus \bf{\hat{a}_{i_n}^{(n)}})}[\ln p(\ten{Y}_{\Omega_T},\ten{S}_{\Omega_T},\{\tilde{\ten{D}}_{\Omega_t}\},\{A^{(n)}\},\lambda,\gamma,\tau)]\\
=&\mathbb{E}_{q}\Bigg[\frac{|\Omega_T|}{2}\ln\tau-\frac{\tau}{2}\left\|\left(\ten{Y}-\ten{A}_T-\ten{S}\right)_{\Omega_T}\right\|_F^2
+\sum_{t=i}^{T-1}\left\{\frac{|\Omega_t|}{2} \ln {(\tau\mu^{T-t})} - \frac{\tau\mu^{T-t}}{2}\left\|\tilde{\ten{D}}_{t}-\ten{A}_t\right\|_F^2\right\}-\frac{1}{2} \hat{\mat{a}}_{i_n}^{(n)}\mat{\Lambda} \hat{\mat{a}}_{i_n}^{(n)T}\Bigg]+\text{const}\\%line 2
=&\mathbb{E}_{q}\Bigg[-\frac{\tau}{2}\left\|\left(\ten{Y}-\ten{A}_T-\ten{S}\right)_{\Omega_T}\right\|_F^2
+\sum_{t=i}^{T-1}\left\{- \frac{\tau\mu^{T-t}}{2}\left\|\left(\tilde{\ten{D}}_{t}-\ten{A}_t\right)_{\Omega_t}\right\|_F^2\right\}-\frac{1}{2} \hat{\mat{a}}_{i_n}^{(n)}\mat{\Lambda} \hat{\mat{a}}_{i_n}^{(n)T}\Bigg]+\text{const}\\%line 3
=&\mathbb{E}_q\Bigg[-\frac{\tau}{2}\left\| \ten{A}_{\Omega_T}\right\|_F^2+\tau\langle\ten{Y}-\ten{S}, \ten{A}_T\rangle_{\Omega_T}
+\sum_{t=i}^{T-1}\left\{ - \frac{\tau\mu^{T-t}}{2}\left\| \ten{A}_{\Omega_t}\right\|_F^2+{\tau\mu^{T-t}}\langle\tilde{\ten{D}}_t, \ten{A}_t\rangle_{\Omega_t} \right\}-\frac{1}{2} \hat{\mat{a}}_{i_n}^{(n)}\mat{\Lambda} \hat{\mat{a}}_{i_n}^{(n)T}\Bigg]+\text{const}\\%line 4
=&\mathbb{E}_q\Bigg[-\frac{\tau}{2}\sum_{t=i}^T\left\{\mu^{T-i}\left\|\ten{A}_{\Omega_t}\right\|_F^2\right\}-\frac{1}{2} \hat{\mat{a}}_{i_n}^{(n)}\mat{\Lambda} \hat{\mat{a}}_{i_n}^{(n)T}+\tau\langle\ten{Y}-\ten{S}, \ten{A}_T\rangle_{\Omega_T}
+\sum_{t=i}^{T-1}\left\{ {\tau\mu^{T-t}}\langle\tilde{\ten{D}}_t, \ten{A}_t\rangle_{\Omega_t} \right\}\Bigg]+\text{const}\\%line 5
=&-\frac{1}{2}\hat{\mat{a}}_{i_n}^{(n)}\Bigg( \mathbb{E}[\tau]\mathbb{E}\left[ \sum_{t=i}^T{\mu^{T-i}}\left(\left(\bigodot_{\substack{k\neq n \\  t-i+1}}\mat{A}^{(k)}\right)_{\Omega_t}^T\left(\bigodot_{\substack{k\neq n \\  t-i+1}}\mat{A}^{(k)}\right)_{\Omega_t}\right)\right] +\mathbb{E}[\mat{\Lambda}]\Bigg)\hat{\mat{a}}_{i_n}^{(n)T}\\%line 11
&+\hat{\mat{a}}_{i_n}^{(n)}\mathbb{E}[\tau]\left(\left(\bigodot_{\substack{k\neq n\\ T-i+1}}\bar{\mat{A}}^{(k)}\right)_{\Omega_T}^T\text{vec}\left(\ten{Y}_{\Omega_T}-\ten{S}_{\Omega_T}\right)+\sum_{t=i}^{T-1} \mu^{T-t}\left(\bigodot_{\substack{k\neq n \\ t-i+1}}\bar{\mat{A}}^{(k)}\right)_{\Omega_t}^T\text{vec}\left(\tilde{\ten{D}}_{\Omega_t,i_n}\right)\right)+\text{const}%line 12
\end{split}
\end{align}
\hrulefill
% The spacer can be tweaked to stop underfull vboxes.
\vspace*{4pt}
\end{figure*}

In (\ref{eq:noise compute}), we further present the derivation of the updates (\ref{eq:update tau}) for the noise precision parameter $\tau$. Here our goal is to identify the parameters of a gamma distribution. By rearranging we obtain the shape-rate parametrization attached to the coefficients $\ln(\tau)$ and $\tau$ respectively. Because our algorithm removes the sparse errors from past tensors the past residuals may not accurately represent the noise $\tau$. Therefore our actual update in (\ref{eq:update tau}) discards past residuals.

\begin{figure*}%[b]
% \vspace*{4pt}
% \hrulefill
\begin{align}\label{eq:noise compute}
\small
\begin{split}
\ln q(\tau) &=\mathbb{E}_{q(\Theta \setminus \tau)}[\ln p(\ten{Y}_{\Omega_T},\ten{S}_{\Omega_T},\{\tilde{\ten{D}}_{\Omega_t}\},\{\mat{A}^{(n)}\},\lambda,\gamma,\tau)]\\\\ 
&=\mathbb{E}_q\Bigg[-\frac{\tau}{2}\left\|\left(\ten{Y}-\ten{S}- \ten{A}_T \right)_{\Omega_T} \right\|_F^2-\sum_{t=i}^{T-1} \frac{\tau\mu^{T-t}}{2}\left\| \left(\tilde{\ten{D}}_t-\ten{A}_t \right)_{\Omega_t} \right\|_F^2+\frac{1}{2}\sum_{t=i}^T\left\{ \ln(\tau\mu^{T-i})|\Omega_t|\right\}-b_0^\tau \tau +(a_0^\tau -1)\ln(\tau)
\Bigg]\\
&=\mathbb{E}_q\left[ \ln(\tau)\left(a_0^\tau -1 +\frac{1}{2}\sum_{t=i}^T|\Omega_t|\right)-\tau\Bigg( b_0^\tau +\frac{1}{2}\left\| \left(\ten{Y}-\ten{S}- \ten{A}_T \right)_{\Omega_T} \right\|_F^2+\sum_{t=i}^{T-1} \frac{\tau\mu^{T-t}}{2}\left\|\left(\tilde{\ten{D}}_t-\ten{A}_t \right)_{\Omega_t} \right\|_F^2   \Bigg)\right]\\
&=\ln(\tau)\Bigg(a_0^\tau -1 +\frac{1}{2}\sum_{t=i}^T|\Omega_t|\Bigg)-\tau\mathbb{E}_q\left[ \Bigg( b_0^\tau +\frac{1}{2}\left\| \left(\ten{Y}-\ten{S}- \ten{A}_T \right)_{\Omega_T} \right\|_F^2+\sum_{t=i}^{T-1} \frac{\mu^{T-t}}{2}\left\|\left(\tilde{\ten{D}}_t-\ten{A}_t \right)_{\Omega_t} \right\|_F^2   \Bigg)\right]
\end{split}
\end{align}
\hrulefill
% The spacer can be tweaked to stop underfull vboxes.
\vspace*{4pt}
\end{figure*}

The remaining updates are similar to those in~\cite{zhao2016bayesian} and can be derived from the computations given in the appendix of~\cite{zhao2016bayesian}. 

\section{Numerical Results}
Our algorithm has been implemented in Matlab. In this section, we verify our algorithm by a synthetic example and several realistic streaming tensor datasets (including surveillance video, dynamic MRI and network traffic). We also compare our proposed method with several existing streaming tensor factorization and completion methods: Online-CP~\cite{online-cp}, Online-SGD~\cite{online-SGD} and OLSTEC \cite{kasai2016online}. The Online-CP and OLSTEC solve essentially the same optimization problem, but Online-CP does not support incomplete tensors. Therefore, our algorithm is only compared with OLSTEC and Online-SGD for the completion task. Our Matlab codes to reproduce all figures and results can be downloaded from \url{www.github.com/anonymous}.
\begin{figure}[t]
\begin{center}
\includegraphics[width=.5\textwidth]{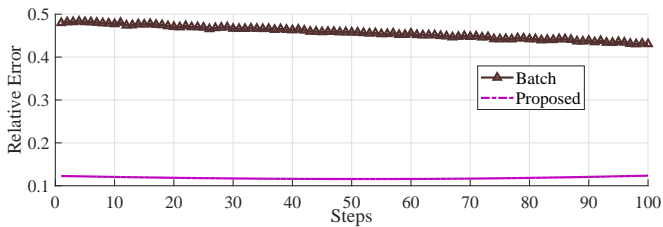}
\caption{Relative Error on a stream of rank-5 $40\times 40$ matrices with $1\%$ of entries sparsely corrupted and $15\%$ of entries sampled.}
\label{fig:batch completion}
\end{center}
\end{figure}

\subsection{Synthetic Data}

\begin{figure*}[t!]
\centering
\subfigure[Factorization for two frames. (top) $10^{th}$ frame in sequence (bottom) $50^{th}$ frame in sequence.]{\label{video_factorization_comparison}\includegraphics[width=.45\textwidth,height=5cm]{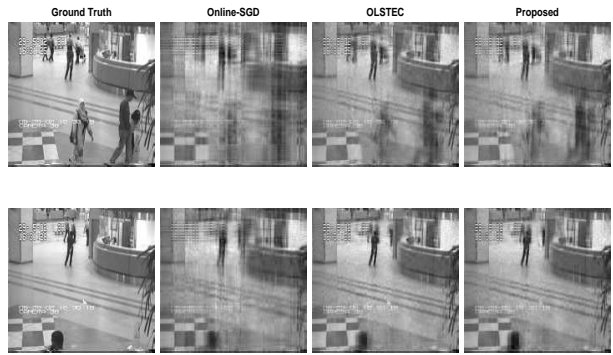}} \hspace{0.2in}
\subfigure[Completion for two frames with $85\%$ missing entries. (top) $10^{th}$ frame in sequence (bottom) $50^{th}$ frame in sequence.]{\label{fig:missing subtraction}\includegraphics[width=.45\textwidth,height=5cm]{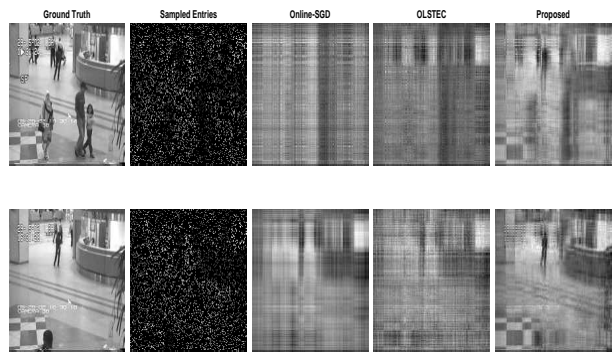}}
\caption{Comparison to existing streaming tensor factorization and completion algorithms on video data. }
\end{figure*}

\begin{figure}[t]
\begin{center}
\includegraphics[width=.5\textwidth]{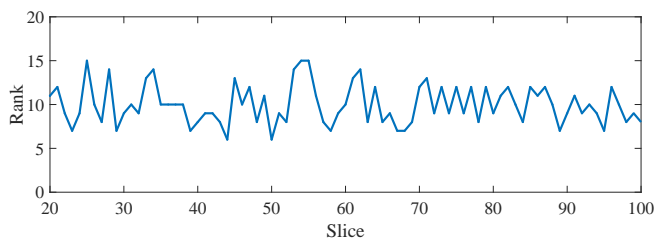}
\caption{Automatically determined rank over time.}
\label{fig:ARD}
\end{center}
\end{figure}

We generate a stream $\{\tilde{\ten{X}}_t\}$ of 100 rank-5 $40\times 40$ matrices. To incorporate temporal drift we randomly generate two sets of factor matrices $\{\mat{P}^{(k)}\}_{k=1}^N$ and $\{\mat{Q}^{(k)}\}_{k=1}^N$ and use a convex combination that changes over time. At time slice $t$ the $k^{th}$ low-rank factor matrix of $\tilde{\ten{X}}_t$ is
\[
\left(1-\frac{t}{100}\right)\mat{P}^{(k)}+\frac{t}{100}\mat{Q}^{(k)}.
\]
The mean entry size of each $\ten{X}_t$ is approximately $1$. Next we generate a stream of sparse error terms $\ten{S}_t$ with $2\%$ nonzero entries of magnitude $10$. We generate our test stream according to assumption (\ref{eq:slice L+S}) by
\begin{equation}\label{eq:slice L+S test data}
\ten{X}_t = \tilde{\ten{X}}_t+\ten{S}_t+\ten{E}_t
\end{equation}
where $\ten{E}_t$ is a dense Gaussian noise term with mean $0$ and variance $10^{-2}$. 

In streaming tensor factorization and completion, we consider the noisy corrupted streaming data $\{\ten{X}_t\}$, and use different numerical methods to recover the hidden factors $\{\mat{A}^{(k)}\}_{k=1}^N$, $ \hat{\mat{a}}_t^{N+1}$ and outliers $\ten{S}_t$ at each time point $t$. We evaluate the accuracy at each time slice based on deviation from the underlying low-rank term $\tilde{\ten{X}}_t$.
\[
\|\tilde{\ten{D}}_t - [\![\mat{A}^{(1)},\dots,\mat{A}^{(N)};\hat{\mat{a}}^{(N+1)}_{t}]\!] \|_{\rm F} /\|\tilde{\ten{D}}_t\|_{\rm F}.
\]
We first compare our method with Online-CP\cite{online-cp}, Online-SGD~\cite{online-SGD} and OLSTEC \cite{kasai2016online} for factoring the full streaming tensor. Then we compare our method only with Online-SGD~\cite{online-SGD} and OLSTEC \cite{kasai2016online} on streaming tensor completion, since Online-CP does not support completion. When factoring the incomplete streaming data, only 15$\%$ randomly sampled data elements are provided. In all methods, the unknown tensor factors are initialized with a maximum rank of $5$. As shown in Fig.~\ref{fig: big synthetic}, our method has better accuracy than all three existing methods for factoring both full and incomplete streaming tensors. Since the sampling set $\Omega_t$ changes as time evolves, any individual sampled slice may have a variable number of outliers. The performance of OLSTEC and Online-SGD highly depends on the number of outliers, and these outliers account for most of reconstruction errors in streaming tensor factorization and completion.

We further compare our streaming factorization method with robust Bayesian CP tensor completion~\cite{zhao2016bayesian}. When testing the method in~\cite{zhao2016bayesian},  we assemble all streaming $\{\ten{X}_t\}$ along the time dimension to create a $40\times 40 \times 100$ tensor. As shown in Fig.~\ref{fig:batch completion}, the Bayesian robust tensor factorization in~\cite{zhao2016bayesian} fails to capture the temporal variation with a good accuracy.

Our final test is to verify the capability of automatic rank determination. We generate a stream $\{\tilde{\ten{X}}_t\}$ of $100$ CP rank-10 $50\times 50$ matrices using the same procedure as above. We generate a stream of sparse error terms $\ten{S}_t$ with $10\%$ nonzero entries of magnitude $10$. We then form a stream sparsely corrupted low-rank tensors as in Equation (\ref{eq:slice L+S test data}). We sample $10\%$ of the entries and run our algorithm to determine the rank. We use a window size of $20$ and the forgetting factor $\mu=0.8$. Despite many sparse corruptions and a small number of samples, our algorithm can adaptively estimate the rank as time evolves. Please note that in streaming tensor completion, we aim to approximate all tensors in a window simultaneously, therefore, the tensor rank is generally larger than the rank of each slice. This is consistent with our result in Fig.~\ref{fig:ARD}.

\subsection{Airport Hall Surveillance Video}
We now test the algorithms on a Airport Hall video data set from the OLSTEC release~\cite{kasai2016online}. In this streaming tensor, each slice is a $144\times 176$ matrix describing a gray-scale video.

Our first task is a low-rank factorization of the full streaming dataset. We set the CP rank to 15. Low-rank factorizations should capture the fixed background despite moving people in the foreground. 
The results for the $10^{th}$ frame and $50^{th}$ frame are both shown in Figure \ref{video_factorization_comparison}. The Online-SGD method~\cite{online-SGD} performs comparably to our method, but requires the full tensor. The OLSTEC method~\cite{kasai2016online} suffers from significant accuracy degradation as time evolves.

We further perform reconstruction of this video sequence using $15\%$ randomly sampled entries. The reconstruction results are shown in Fig.~\ref{fig:missing subtraction}. On this task our algorithm outperforms both OLSTEC~\cite{kasai2016online} and Online-SGD~\cite{online-SGD} due to its capability of capturing the underlying sparse outliers.

\subsection{Dynamic Cardiac MRI}

\begin{figure}[t]
\begin{center}
\includegraphics[width=.45\textwidth,height=5cm]{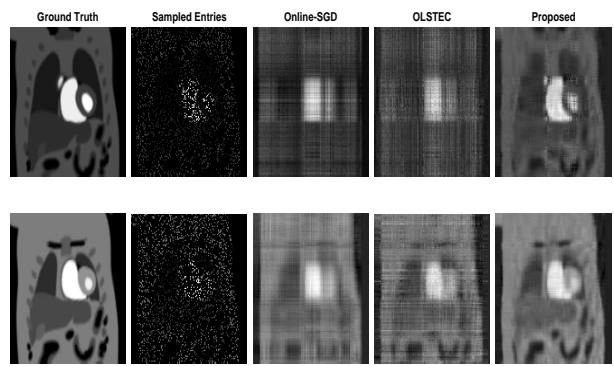}
\caption{MRI reconstruction via streaming tensor completion.}
\label{fig:mri reconstruction}
\end{center}
\end{figure}

Next we consider a dynamic cardiac MRI dataset from \cite{sharif2007physiologically} and obtained via \url{https://statweb.stanford.edu/~candes/SURE/data.html}. Each slice of this streaming tensor dataset is a $128\times 128$ matrix. In clinical applications, it is highly desirable to reduce the number of MRI scans. Therefore, we are interested in using streaming tensor completion to reconstruct the whole sequence of medical images based on a few sampled entries. The underlying structure of the cardiac muscle remains fixed over time but heartbeats introduce contractions that make a low-rank completion difficult.

In all methods we set the underlying maximum rank to $15$. For our algorithm we set the forgetting factor to $\mu = 0.98$ and the the sliding window size to $20$. In OLSTEC we set the forgetting factor to the suggested default of $0.7$ and the sliding window size to $20$. The available implementation of Online-SGD does not admit a sliding window, but instead computes with the full (non-streamed) tensor. While this may limit its ability to work with large streamed data in practice, we include it in comparison for completeness. With $15\%$ random samples, the reconstruction results are shown in Fig.~\ref{fig:mri reconstruction}. The ability of our model to capture both small-magnitude measurement noise and sparse large-magnitude deviations renders it more effective than OLSTEC and Online-SGD for this dynamic MRI reconstruction task.

\subsection{Multimodal Dynamic Cardiac MRI}

\begin{figure}[t]
\begin{center}
\includegraphics[width=.45\textwidth,height=5cm]{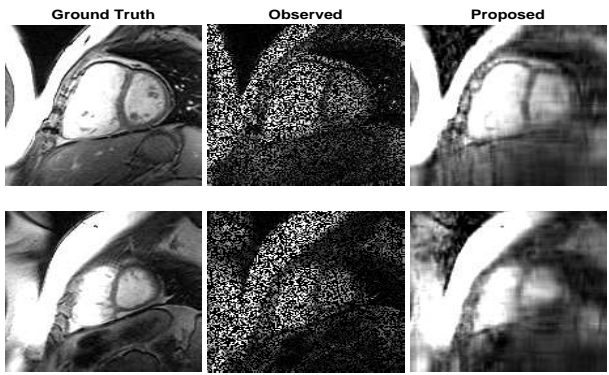}
\caption{Multimodal MRI reconstruction via streaming tensor completion.}
\label{fig:multimodal mri reconstruction}
\end{center}
\end{figure}

We further test our algorithm on a higher-dimensional cardiac MRI dataset from~\cite{andreopoulos2008efficient}. Each temporal slice is a 3D tensor of size $150\times 150\times 5$ that describes the entire cardiac muscle rather than a 2D cross-section. We set the maximum rank to $30$, the forgetting factor to $\mu=0.98$, and the sliding window size to $5$. Since our method is the only one capable of handling higher-order tensor completion, only the results of our algorithm are shown. The reconstruction results are shown for $50\%$ missing samples in Figure \ref{fig:multimodal mri reconstruction}. We display the results of our algorithm from two cross sections obtained at the same time point. 

\subsection{Network Traffic}
Our final example is the Abilene network traffic dataset \cite{lakhina2004structural}. This dataset consists of aggregate Internet traffic between 11 nodes, measured at five-minute intervals. On this dataset we test our algorithm for both reconstruction and completion. The goal is to identify normally evolving network traffic patterns between nodes. If one captures the underlying low-rank structure, one can identify anomalies for further inspection. Anomalies can range from malicious distributed denial of service (DDoS) attacks to non-threatening network traffic spikes related to online entertainment releases. In order to classify abnormal behavior one must first fit the existing data.
We evaluate the accuracy of the models under comparison by calculating the relative prediction error at each time slice:
\[
\|\ten{X}_t - [\![\mat{A}^{(1)},\dots,\mat{A}^{(N)},\hat{\mat{a}}^{(N+1)}_{t}]\!] - \ten{S}_t\|_{\rm F} /\|\ten{X}_t\|_{\rm F}.
\]

We provide a comparison of different methods on the full dataset in Fig.~\ref{fig:network factorization}. In order to provide a realistic setting we exclude a ``burn-in" time of 10 frames, after which the error patterns are stable. Our algorithm significantly outperforms OLSTEC and Online-SGD in factoring the whole data set.
\begin{figure}[t]
\begin{center}
\includegraphics[width=.49\textwidth]{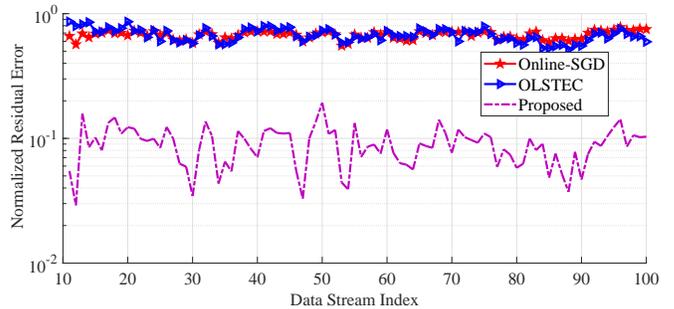}
\caption{Factorization error of network traffic from complete samples.}
\label{fig:network factorization}
\end{center}
\end{figure}

Then we remove $50\%$ of the entries from the the Abilene tensor and attempt to reconstruct the whole network traffic. Our results are shown in Fig.~\ref{fig:network completion}. Again we use a ``burn-in" time of 10 frames. Unlike the MRI and video data examples, this is an example in which the streaming data size is relatively small ($11\times 11$) and therefore we may not require all 15 available rank-1 factors. Since existing streaming tensor completion algorithms assume a fixed-rank, they are likely to either over-fit or under-fit the data. The adaptive rank selection of our algorithm avoids both drawbacks.

\begin{figure}
\begin{center}
\includegraphics[width=.49\textwidth]{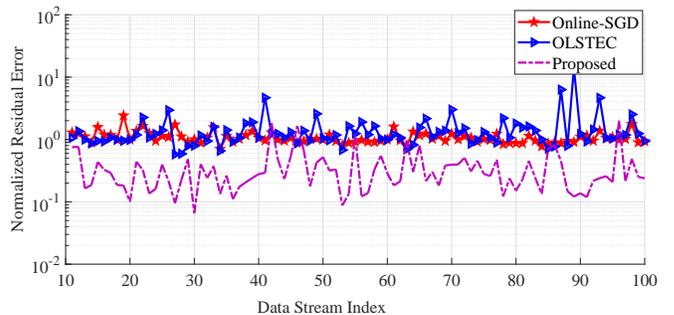}
\caption{Reconstruction error of network traffic with $50\%$ of data missing.}
\label{fig:network completion}
\end{center}
\end{figure}
% \begin{figure}
% \begin{center}
% \includegraphics[width=.45\textwidth,height=5cm]{adaptive_rank_selection}
% \caption{Adaptive rank selection for network traffic data completion.}
% \label{fig:adaptive rank}
% \end{center}
% \end{figure}

\section{Conclusion}

We have presented a probabilistic model for low-rank plus sparse streaming tensor factorization and completion. We have proposed a variational Bayesian solver and tested our solver on both real and synthetic data. We have demonstrated the performance of our algorithm for tensor data applications in dynamic MRI, network traffic monitoring, and video surveillance. Our algorithm outperforms existing approaches due to their reliance on a least-squares cost function that is vulnerable to outliers. We have also shown that our algorithm avoids over-fitting by automatically determining the rank. %Our work provides the extension of robust tensor methods to the streaming problem and demonstrates the practical advantages of this new approach.

\bibliographystyle{IEEEtran}
\bibliography{egbib}
%}

\newpage

\end{document}

%% file: DefinedComand.tex
\newcommand{\normalpdf}[3]{\mathcal{N}\left(#1 \middle\vert#2 , #3\right)}

\newcommand{\hadamard}{\operatornamewithlimits{\mathlarger{\mathlarger{\mathlarger{\circledast}}}}}

\theoremstyle{plain}
\newtheorem{theorem}{Theorem}
\newtheorem{lemma}{Lemma}

\theoremstyle{definition}
\newtheorem{definition}[theorem]{Definition}
\theoremstyle{remark}